\documentclass[aps,prl,reprint,superscriptaddress]{revtex4-1}
\usepackage{amsmath,graphicx}

\begin{document}

\title{Mechanically Detecting and Avoiding the Quantum Fluctuations\\ of a Microwave Field}

\author{J. Suh}
\affiliation{Applied Physics, Caltech, Pasadena, CA,  91125 USA}
\affiliation{Kavli Nanoscience Institute, Caltech, Pasadena, CA 91125 USA}
\author{A. J. Weinstein}
\affiliation{Applied Physics, Caltech, Pasadena, CA,  91125 USA}
\affiliation{Kavli Nanoscience Institute, Caltech, Pasadena, CA 91125 USA}
\author{C. U. Lei}
\affiliation{Applied Physics, Caltech, Pasadena, CA,  91125 USA}
\affiliation{Kavli Nanoscience Institute, Caltech, Pasadena, CA 91125 USA}
\author{E. E. Wollman}
\affiliation{Applied Physics, Caltech, Pasadena, CA,  91125 USA}
\affiliation{Kavli Nanoscience Institute, Caltech, Pasadena, CA 91125 USA}
\author{\\S. K. Steinke}
\affiliation{Applied Physics, Caltech, Pasadena, CA,  91125 USA}
\affiliation{Department of Physics, University of Arizona, Tuscon, AZ 85721 USA}
\author{P. Meystre}
\affiliation{Department of Physics, University of Arizona, Tuscon, AZ 85721 USA}
\author{A. A. Clerk}
\affiliation{Department of Physics, McGill University, Montreal, QC, H3A 2T8 CA}
\author{K. C. Schwab}
\thanks{Correspondence and requests for materials should be addressed to Keith Schwab~(email: schwab@caltech.edu).}
\affiliation{Applied Physics, Caltech, Pasadena, CA,  91125 USA}
\affiliation{Kavli Nanoscience Institute, Caltech, Pasadena, CA 91125 USA}

\begin{abstract}
During the theoretical investigation of the ultimate sensitivity of gravitational wave detectors through the 1970's and '80's, it was debated whether quantum fluctuations of the light field used for detection, also known as photon shot noise, would ultimately produce a force noise which would disturb the detector and limit the sensitivity.  Carlton Caves famously answered this question with ``They do.'' \cite{Caves:1980PRL}  With this understanding came ideas how to avoid this limitation by giving up complete knowledge of the detector's motion\cite{Braginskii:1975,Thorne:1978,Braginsky:1980}.  In these back-action evading (BAE) or quantum non-demolition (QND) schemes, one manipulates the required quantum measurement back-action by placing it into a component of the motion which is unobserved and dynamically isolated.  Using a superconducting, electro-mechanical device, we realize a sensitive measurement of a single motional quadrature with imprecision below the zero-point fluctuations of motion, detect both the classical and quantum measurement back-action, and demonstrate BAE avoiding the quantum back-action from the microwave photons by 9 dB. Further improvements of these techniques are expected to provide a practical route to manipulate and prepare a squeezed state of motion with mechanical fluctuations below the quantum zero-point level, which is of interest both fundamentally\cite{Hu:1993} and for the detection of very weak forces\cite{Clerk:2008}.
\end{abstract}

\maketitle

Since the discovery of Shor's Algorithm\cite{Shor:1997} almost 20 years ago, a major theme in physics has been about the untapped power and benefits of quantum phenomena, largely stemming from the resource of quantum entanglement.  However much earlier, it was understood how quantum physics places limits on our knowledge\cite{Drever:1976,Caves:1980RMP}.  This limitation can be useful, as in the case of quantum cryptography schemes where the required quantum measurement back-action of an eavesdropper leaves its trace on the transmitted information, providing proof of their snooping.  For measurements of position, this limitation, called the Standard Quantum Limit (SQL)\cite{Caves:1980RMP} is not beneficial:  back-action due to the quantum nature of the measurement field, ultimately obscures our vision for a sufficiently sensitive measurement.

Quantum limitations on the detection of position are no longer academic issues; in recent years, the detection of motion has now advanced to the point that quantum back-action engineering is now required to improve the sensitivity. Detections of motion have been realized with imprecision below that at SQL\cite{Tuefel:2009,Anetsberger:2010}. Back-action forces from the quantum noise of the detection field have been demonstrated to drive the motion of mechanical oscillators, first with electrons in an electro-mechanical structure\cite{Naik:2006} and then with photons in opto-mechanical systems \cite{Murch:2008,Purdy:2013}. In this work, we demonstrate the back-action forces due to the shot noise of microwave photons, which are $10^4$ times lower in energy than optical photons.

Strategies to manipulate the quantum measurement back-action have included modifying the quantum fluctuations of the measurement field\cite{Caves:1981,LIGO:2011}, and modulating the coupling between the detection field and mechanical element. The modulation of coupling can be implemented by either sudden stroboscopic measurement\cite{Braginsky:quantum,Ruskov:2005} or continuous two-tone BAE measurement\cite{Braginskii:1975,Thorne:1978,Braginsky:1980,Clerk:2008}, which we pursue here.    

The system we study is a parametric transducer (Fig. 1a): a mechanical resonator ($\omega_m=2\pi\cdot$4.0 MHz) modulates the capacitance of a superconducting electrical resonator ($\omega_c=2\pi\cdot$5.4 GHz), and modifies $\omega_c$ by 14 Hz ($=g_0/2\pi$) per $x_{zp}$, where $x_{zp}=\sqrt{\hbar/(2m\omega_m)}\approx 1.8$ fm is the amplitude of zero-point fluctuations of the mechanical resonator with mass $m$. The damping rate of the electrical resonator is $\kappa=2\pi\cdot$0.86 MHz, which places this system into the side-band resolved limit ($\omega_m>\kappa$) required to realize BAE dynamics\cite{Bocko:1996,Clerk:2008}. When pumping the transducer with microwave photons at $\omega_p=\omega_c-\omega_m$, the electro-mechanical coupling together with mechanical motion results in frequency up-conversion of pump photons to $\omega_c$ in a Raman-like process at a rate $n_m\Gamma_{opt}$, where $n_m$ is the occupation factor of the mechanical mode, and $\Gamma_{opt}=4g_0^2 n_p / \kappa$ with $n_p$ as the number of pump photons stored in the electrical resonator. Similarly, when pumping at $\omega_p=\omega_c+\omega_m$, photons are down-converted at a rate $(n_m+1)\Gamma_{opt}$. These sidebands are the signals we analyze in this work: we use thermal motion of the mechanical resonator at calibrated temperatures to measure the transduction gain between the sideband power and $n_m$\cite{Hertzberg:2010}. Based upon the thermal calibration, we observe a motional sideband corresponding to 7.2$\pm$0.2 mK at the base temperature of our refrigerator (Fig.1c, red cross and inset). The thermal calibration also determines $g_0$, and we combine it with back-action damping measurement\cite{Marquardt:2007,Rocheleau:2010}, to generate the calibration of $n_p$ vs measured microwave power (Fig.1d).

\begin{figure}
\centering\includegraphics{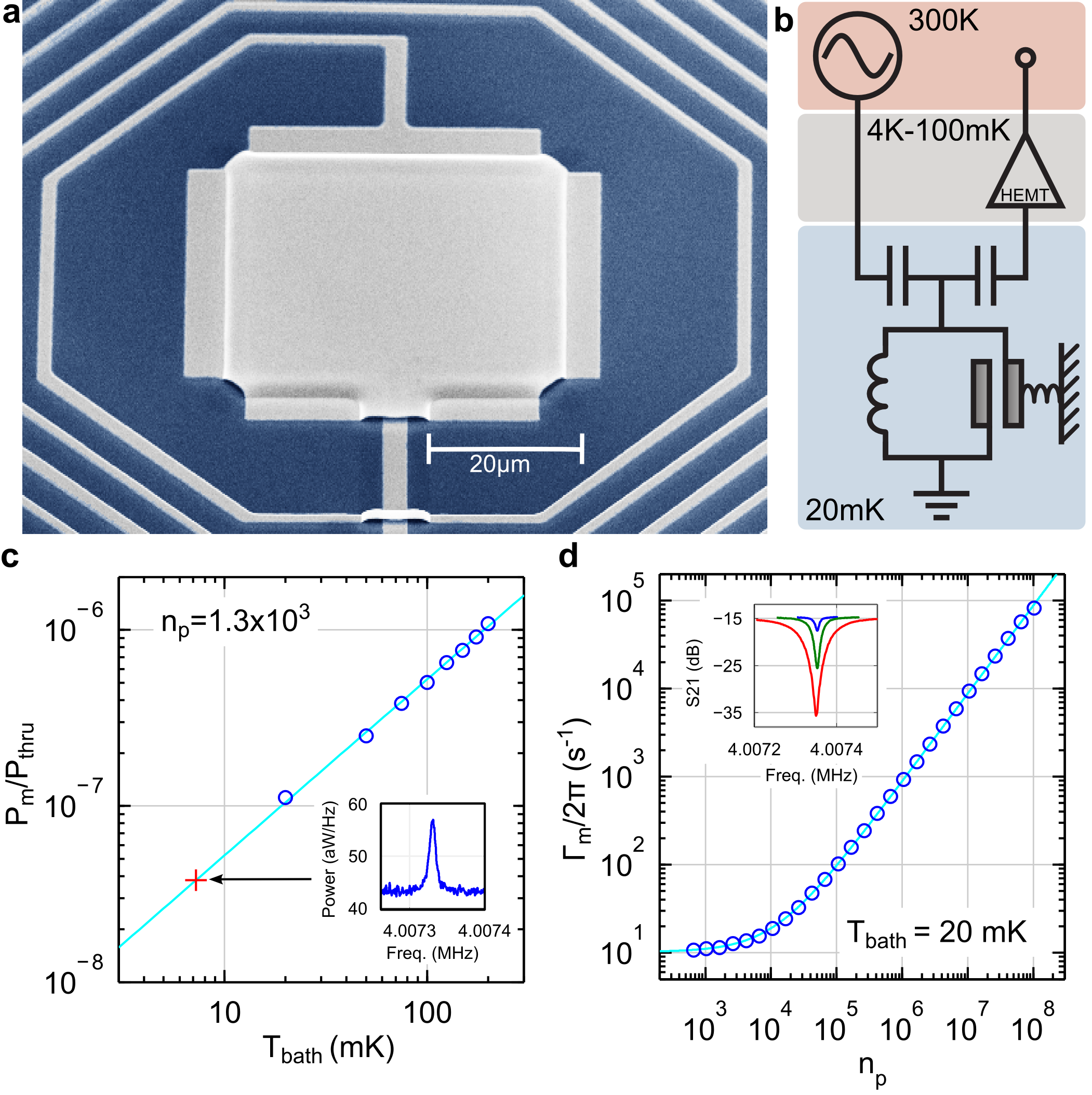}
\caption{Device, measurement scheme and sample characterization. \textbf{a}, Device electron micrograph. A parallel plate capacitor is connected to a spiral inductor, forming a lumped-element microwave resonator. The top plate of the capacitor is a compliant membrane, and we study its fundamental mechanical resonance here. Color indicates different materials used: blue - silicon, gray - aluminum. \textbf{b}, Measurement scheme. Shot-noise-limited microwave tones are generated with room-temperature filtering and cryogenic attenuation, and applied to the device at 20 mK. The output microwave field from the device is amplified by a low-noise amplifier at 4.2K, and its spectra are analyzed (Supplementary Information). \textbf{c}, Calibration of motional sidebands against thermal motion (blue circles and line). Inset, motional sideband at base temperature corresponding to $7.2\pm0.2$ mK. \textbf{d}, Back-action damping and calibration for the number of pump photons. In addition to a red-detuned pump, a weak probe sweeping near the electrical resonance is applied and its absorption is monitored, showing the resonant mechanical response. Blue circle, mechanical damping rate. Blue line, back-action damping theory fit\cite{Marquardt:2007}. Inset, examples of absorption spectra at $n_p\approx5\times10^3,3\times10^4,1\times10^5$ from top to bottom.}
\end{figure}

Two-tone BAE is accomplished in this system with the application of a modulated electric field: $E(t)=E_p\cos\omega_mt\cdot\cos\omega_ct=\left(E_p/2\right)[\cos(\omega_c-\omega_m)t+\cos(\omega_c+\omega_m)t]$. This technique has the effect of coupling to a single mechanical quadrature $X_1$, where $x(t)=X_1(t)\cos\omega_mt+X_2(t)\sin\omega_mt$. The BAE nature can be understood by noting that the back-action force produced at the mechanical frequency, by the beating between voltage noise at the microwave resonance and the pump field, produces displacements exclusively in the $X_2$ quadrature\cite{Bocko:1996}. In this way, one gains information about $X_1$ and places the required quantum measurement back-action into $X_2$.  This fact can be also understood by observing that $X_1$ and $X_2$ are constants of motion and thus quantum non-demolition observables\cite{Thorne:1978}.

Recent experiments have attempted to implement two-tone BAE demonstrating a single quadrature imprecision close to the zero-point fluctuations, with further sensitivity improvement blocked by parametric instabilities arising from non-linearities of the coupling\cite{Hertzberg:2010}, thermal dissipation\cite{Suh:2012}, and two-level system effects\cite{Suh:2013}. In the experiments described here, we study an improved device which largely avoids these limitations. However, to overcome difficulties in aligning BAE tones arising from a small mechanical frequency jitter comparable to the mechanical damping rate of $\Gamma_{m0}=2\pi\cdot$10 Hz at 20 mK, we also apply a red-detuned cooling tone at $\omega_c-\omega_m-35$kHz to both cool the resonator from occupation factor $n_m\approx100$ to 15, and to broaden the mechanical resonance from $\Gamma_m\sim2\pi\cdot$10 Hz to $2\pi\cdot$ 100Hz\cite{Marquardt:2007,Rocheleau:2010}. Since this cooling tone is well-separated from the measurement tones with respect to the mechanical linewidth ($\Gamma_m\ll35$ kHz), it effectively adds mechanical damping without perturbing our BAE measurements (Fig.3a).

To demonstrate the avoidance of measurement back-action, we first pump the transducer with two tones of equal power with frequencies of $\omega_c\pm\left(\omega_m+\Delta\right)$, where each tone is detuned by $\Delta = 5$kHz from the BAE configuration, producing two separate motional sidebands.  When $\Delta\gg\Gamma_m$, the up- and down- converted signal photons provide a measurement of both mechanical quadratures, and as a result, the measurement is subject to the usual back-action forces resulting in extra position fluctuations, $\left<x^2\right>_{ba}/x_{zp}^2=2(\Gamma_{opt}/\Gamma_m)(2n_c+1)$, where $n_c$ is the occupation factor of the electrical resonator, and $+1$ is due to the quantum fluctuations of the microwave field. The observed motional sidebands exhibit slight asymmetry mainly due to the interference between microwave noise and mechanical motion\cite{Monroe:1995,Safavi:2012}; we take the average weight of the two sidebands to cancel this effect and extract $\left<x\right>^2$ (Supplementary Information). The blue circles and dots in Fig.2b show the imprecision and back-action versus $n_p$ of each tone: as the imprecision decreases, the fluctuations due to back-action increase, increasing the mechanical occupation from 15 to 65 at $n_p=2.3\times10^6$. In addition to the back-action, the down-converted sideband has 5\% more power than the up-converted one (Fig.2a), which is consistent with the expected asymmetry, and will be the subject of future work. Both the back-action and the asymmetry observed at $n_p=2.3\times10^6$ are consistent with small finite microwave occupation factor ($n_c\approx0.6\pm0.1$) in addition to the quantum fluctuations.

In the BAE configuration ($\Delta=0$), the motional sidebands overlap to become a single peak, and the response is dramatically different. As the imprecision decreases, we do not observe a large increase in the mechanical fluctuations, as shown red dots in Fig.2b with $n_p$ from both tones.  The expected back-action into the measured quadrature due to the finite sideband resolution is\cite{Clerk:2008}: $\left<X_1^2\right>_{ba}=\frac{1}{32}(\frac{\kappa}{\omega_m})^2\left<X_2^2\right>_{ba}\approx\left<X_2^2\right>_{ba}/700\approx 0.12x_{zp}^2$ at $n_p=4.7\times10^6$.  The measured back-action of $\left<X_1^2\right>_{ba}\approx10 x_{zp}^2$ is likely due to ohmic heating of our device.  Nonetheless,  we demonstrate avoidance of the back-action noise by 10 dB at $n_p=2.3\times10^6$.  Most importantly, we show that the back-action $\left<X_1^2\right>_{ba}$ is 9 dB below the level set by quantum fluctuations of the microwave field, $2(\Gamma_{opt}/\Gamma_m)x_{zp}^2$ at $n_p=4.7\times10^6$. The quadrature imprecision is also below $x_{zp}$ at this point: $\left<X_1^2\right>_{imp}\approx0.6x_{zp}^2$ (Fig.2b, inset).

\begin{figure}
\centering\includegraphics{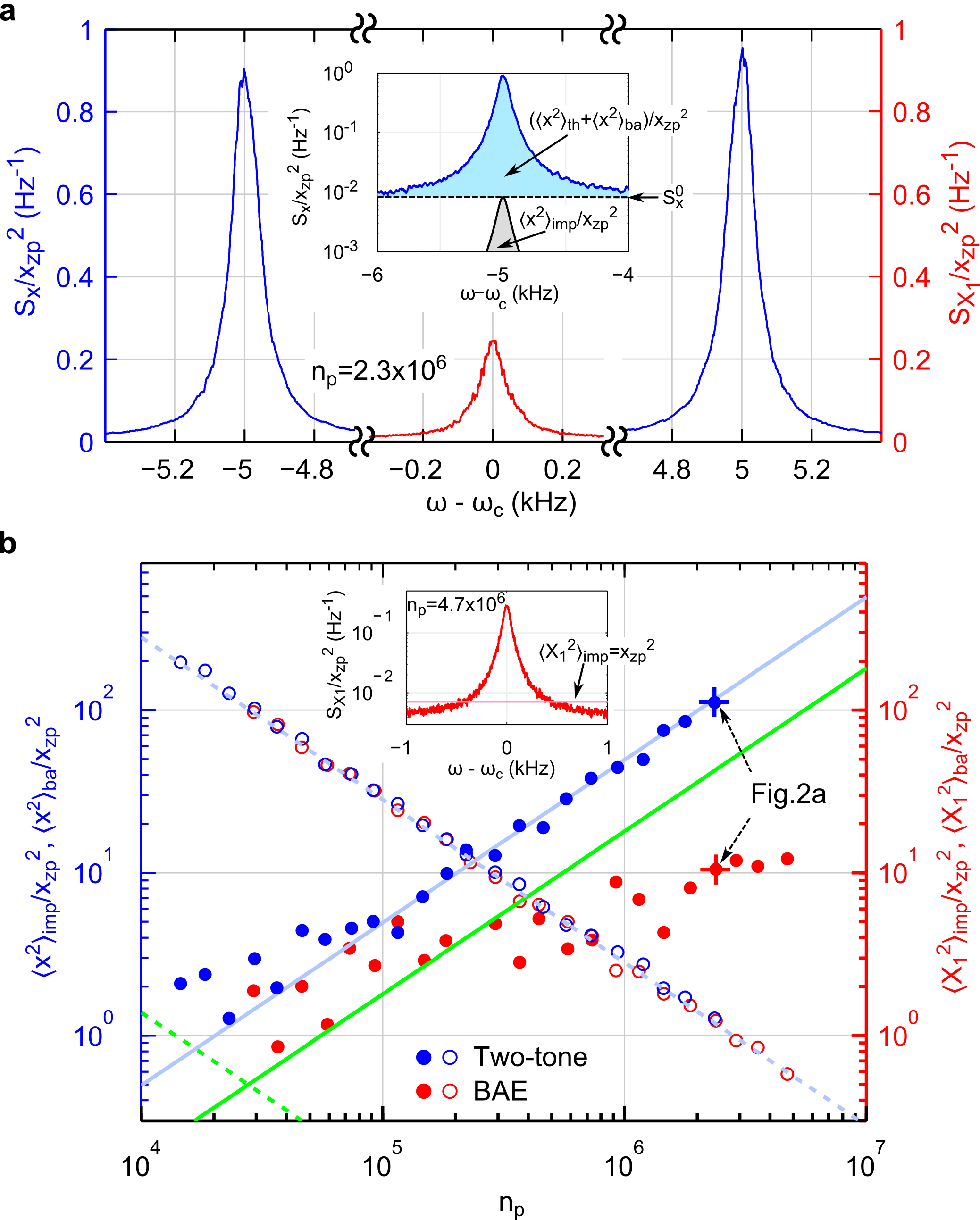}
\caption{Back-action evading measurement. \textbf{a}, An example of measured motional sidebands in two-tone non-BAE (blue) and BAE (red) configurations. The spectra clearly show reduced back-action in the BAE set-up. Inset, the blue area under the Lorentzian peak of the motional sideband, which is either the average of two sidebands in two-tone non-BAE or the single sideband in BAE, represents the fluctuations due to thermal ($\left<x^2\right>_{th}$)and measurement back-action forces $\left<x^2\right>_{ba}$ (Supplementray Information). Imprecision ($\left<x^2\right>_{imp}$), which is the additive noise inferred from the measurement noise floor($S_x^0$), is defined as the gray area under a Lorentzian with its peak at $S_x^0$, and with linewidth $\Gamma_m$. \textbf{b}, Measurement imprecision(circles) and back-action(dots) of two-tone non-BAE(blue) and BAE(red). The solid blue line represents a fit to the measured back-action including classical noise in the electrical resonator. The solid green line is the expected quantum back-action from microwave shot noise. The measured back-action in BAE lies below the quantum back-action above $n_p\approx10^6$. The dashed blue line shows a fit to the measured imprecision and the dashed green line is the imprecision expected at the quantum limit ($=1/(8\Gamma_{opt}/\Gamma_m)$).  Inset, the BAE imprecision reaches $0.6x_{zp}^2$ at $n_p=4.7\times10^6$.}
\end{figure}

Even though our detection amplifier is far from quantum-limited with a noise temperature of about 4 K, the strong BAE measurement results in a detector noise product $\sqrt{S_{X_1}S_F}$ of approximately 2.5$\hbar$, lower than other detection approaches with micro- and nano-mechanical devices\cite{Clerk:2010}. By minimizing device heating due to circuit loss, we would expect this figure to drop to $\sqrt{S_{X_1}S_F}\approx0.3\hbar$, exceeding what is possible for a perfect measurement of position\cite{Clerk:2010}: $\sqrt{S_xS_F}=\hbar/2$. With a nearly quantum-limited amplifier\cite{Mutus:2013}, we expect even further improvement reaching $\sqrt{S_{X_1}S_F}\approx0.08\hbar$.

To show that the avoided back-action is indeed added to the unobserved quadrature, we place a second set of probe BAE tones, 20dB weaker than the pump BAE tones at $n_p=1.1\times10^6$ (Fig.3a). We measure and control the relative phase($\phi$) of these tones, and measure the resulting motional sidebands (Supplementary Information).  The signal from the probes measures a quadrature variance along a rotated axis: $\left<X(\phi)^2\right>=\left<X_1^2\right>\cos^2\phi+\left<X_2^2\right>\sin^2\phi$. Figures 3b-c compare the signals from the two sets of BAE tones.  As is apparent, the fluctuations at $\phi=\pi/2$, aligned along the $X_2$ quadrature, show maximal back-action heating.  

We study the back-action noise into the $X_2$ quadrature ($\left<X_2^2\right>_{ba}$) versus microwave occupation factor by applying microwave noise to increase $n_c$. The HEMT amplifier noise floor is carefully measured using a cryogenic microwave switch, and this is the noise floor used in measuring the increase in noise power $\Delta\eta$ in the electrical resonator due to $n_c$(Supplementary Information). Figure 3d shows the observed $\left<X_2^2\right>_{ba}$ versus the increase in the measured microwave noise power, where $\left<X_2^2\right>_{ba}/x_{zp}^2=2(\Gamma_{opt}/\Gamma_m)(2n_c+1)$ is predicted\cite{Clerk:2008}.  The figure shows good fit to this formula, and since measured microwave power is proportional to $n_c$, the slope of the fit provides a calibration for $n_c$, $n_c/\Delta\eta=0.22\pm0.02$ $\textrm{(aW/Hz)}^{-1}$.
 
Most importantly, the fit intercept shown in Fig.3d of $1.1\pm0.1$ shows excellent agreement to $+1$ expected from quantum back-action, with the contribution of thermal $n_c\approx0.1$ (Supplementary Information). In this way, we show that the mechanical device detects the quantum fluctuations of the microwave field\cite{Koch:1982,Fragner:2008}.

\begin{figure}
\centering\includegraphics{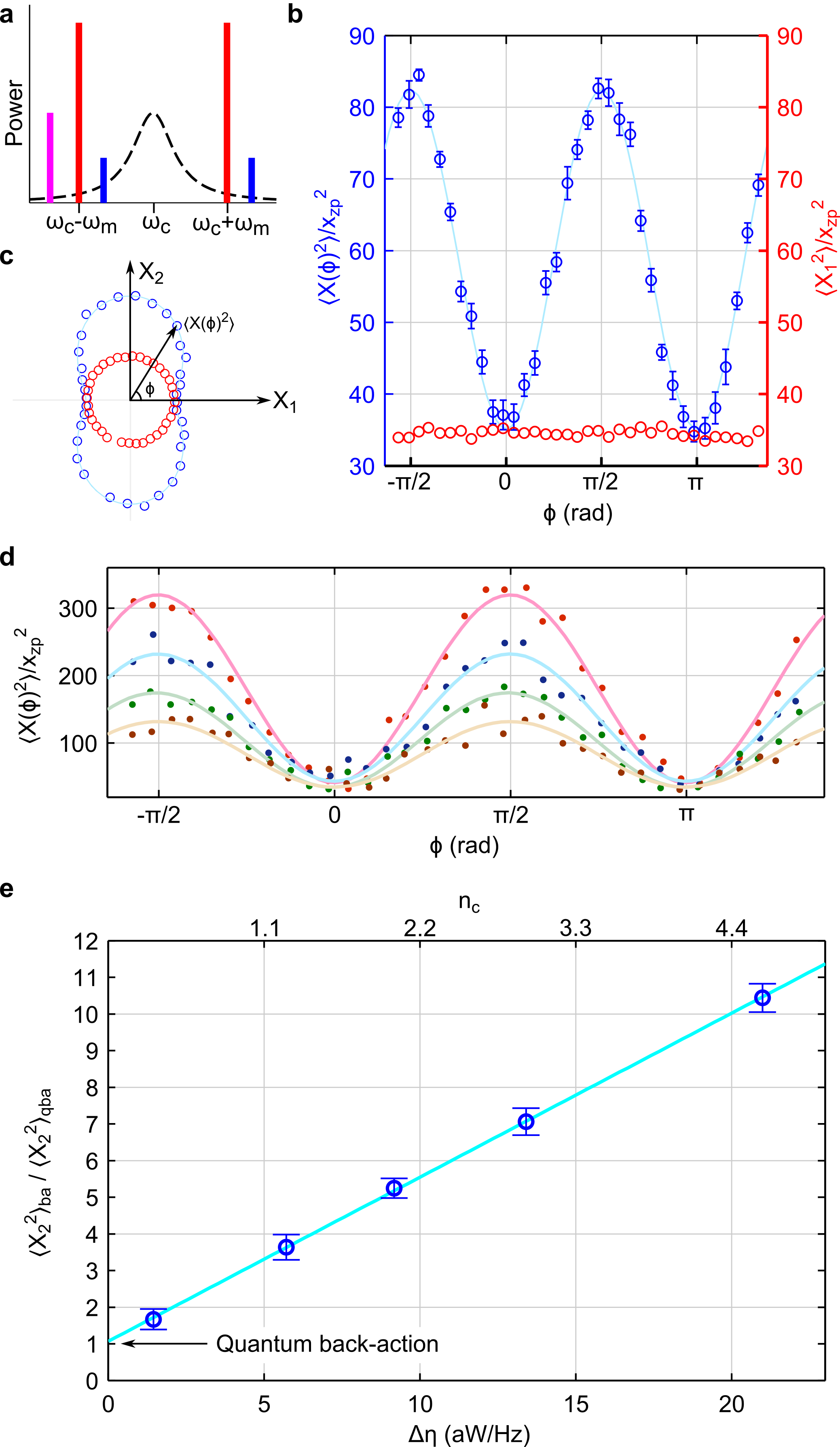}
\caption{Measurement of back-action. \textbf{a}, Microwave set-up. Strong BAE pumps(red) are applied symmetrically about the microwave resonance (dashed line). Weak BAE probes(blue) with a small detuning(+30 kHz) from the pumps are used to measure the back-action from the BAE pumps. A weak cooling tone with $\Gamma_{opt}\approx$90 Hz is applied at the same time (magenta). \textbf{b}, An example of measured mechanical fluctuations along the BAE pump axis (red circles) and the probe axis (blue circles). $\phi$ is the angle between these two axes. The blue line is a fit to $A\sin^2\phi+B$. (Supplementary Information). \textbf{c}, Polar plot of \textbf{b}. It defines $X_1$ and $X_2$ quadratures along the direction of minimum and maximum fluctuation, respectively. \textbf{d}, Mechanical fluctuations along the probe axis at different microwave noise powers: $\Delta\eta=5.7, 9.2, 13, 21$ aW/Hz (brown, green, blue, and red dots, respectively). \textbf{e}, Back-action in the $X_2$ quadrature normalized by quantum back-action $\left<X_2^2\right>_{qba}=2(\Gamma_{opt}/\Gamma_m)x_{zp}^2$.}
\end{figure}

These results lead the way towards manipulating the quantum noise of a mechanical resonator.  As described 
in Ref.~\onlinecite{Clerk:2008}, in a particular run of the experiment, the motion is expected to be highly squeezed: we estimate the conditional squeezing $\left<X_1^2\right> / \left<X_2^2\right>\approx0.01$.  However, after averaging over many runs from a thermal state, we recover the thermal distribution on both quadratures and lose squeezing.  To produce a squeezed state from a thermal state, feedback on the motion may be applied.  Using a nearly quantum limited amplifier\cite{Mutus:2013}, we expect to produce a squeezed state ($\left<X_1^2\right>/ x_{zp}^2 < 1$) with $n_p\approx10^5$, which can be useful for detection of weak forces and fundamental studies of quantum decoherence\cite{Hu:1993}. We also note that the mechanical mode reaching 7.2 mK demonstrates a new application of a micro-mechanical resonator as a primary ultra-low-temperature thermometer.

\begin{acknowledgements}
We would like to acknowledge Jared Hertzberg, Tristan Rocheleau, Tchefor Ndukum, and Matt Shaw for work on earlier experiments which lead to these results. This work is supported by funding provided by the Institute for Quantum Information and Matter, an NSF Physics Frontiers Center with support of the Gordon and Betty Moore Foundation (NSF-IQIM 1125565), by DARPA (DARPA-QUANTUM HR0011-10-1-0066), and by NSF (NSF-DMR 1052647, NSF-EEC 0832819).
\end{acknowledgements}

\end{document}


\title{Supplementary Information for ``Mechanically Detecting and Avoiding the Quantum Fluctuations of a Microwave Field''}
\author{J. Suh, A. J. Weinstein, C. U. Lei, E. E. Wollman, S. K. Steinke, \\
P. Meystre, A. A. Clerk \& K. C. Schwab}        
\date{}        
\maketitle

\section{Sample fabrication}

We start with a 525 $\mu$m thick $<$100$>$-oriented high resistivity ($>$10k$\Omega \cdot$ cm) silicon wafer.  After initial surface preparation, a 100nm layer of aluminum is DC magnetron sputtered in a UHV chamber with base pressure of $\sim10^{-9}$ Torr.  The bottom layer is patterned via contact photolithography followed by two-step wet etching in Transene Al Etchant A and MF319.  Next, we spin and pattern S1813 which acts as a sacrificial layer in the capacitor gap and a protection layer for the rest of the bottom layer pattern.  In order to thin down the sacrificial layer, we flood expose the S1813 prior to development.  Before processing the top aluminum layer, we use a short O$_2$ plasma etch to increase adhesion between the sacrificial and top layer.  A 150nm aluminum layer is sputtered and patterned under the same procedure for the bottom layer. The resulting device is cleaned and released in an overnight soak in Remover-PG followed by critical point drying and a final short O$_2$ plasma clean.

\section{Theory}\label{theory}

We start with a standard linearly coupled optomechanical Hamiltonian,
\begin{equation}
\label{hamiltonian.theory}
\hat H = \hbar(\omega_c - g_0 \hat x/x_{zp})\left [\hat a^\dagger \hat a - \langle \hat a^\dagger \hat a\rangle(t)\right ]+\hbar \omega_m \hat c^\dagger \hat c \nonumber \\
+i\hbar\sqrt{\kappa_L}(\bar{a}^*_{in}(t)\hat a-\bar{a}_{in}(t)\hat a^\dag),
\end{equation}
where $\hat x = x_{zp}(\hat c + \hat c^\dagger)$ is the mechanical position, $\hat c$ is the phonon annihilation operator, $\hat a$ is the photon annihilation operator, $\langle \hat a^\dagger \hat a\rangle(t)$ is the classical mean photon number at time $t$,  $\bar{a}_{in}(t)$ is the driving field, and all other constants are as defined in the text.

We make unitary transformations to a rotating, displaced frame, so that $\hat a = e^{-i\omega_ct}(\alpha(t)+\hat d)$, with $\alpha(t) = \langle \hat a e^{i\omega_c t}\rangle$, and so that $\hat b = \hat c e^{i\omega_mt}$. Including environmental dissipation, we arrive at the linearized Heisenberg equations of motion,
\begin{align}
	\label{heisenberg.theory1}
	\dot {\hat d} &= - i g_0\alpha(t)\left(\hat b e^{-i\omega_m t} + \hat b^\dagger e^{i\omega_m t}\right)-\frac{\kappa}{2}\hat d - 
		\sum_{j = L, R, int} \sqrt{\kappa_j}	e^{i\omega_c t} \hat \xi_j(t) \\
	\label{heisenberg.theory2}
	\dot{\hat b} &= - ig_0 \hat b \left(\alpha(t)\hat d^\dagger + \alpha^*(t)\hat d \right) -\frac{\Gamma_{m0}}{2}\hat b - \sqrt{\Gamma_{m0}}e^{i\omega_m t}\hat \eta(t),
\end{align}
Here, the index $j$ labels the three ports coupled to the cavity:  the right and left coupling ports ($j=R, L$) and a port representing internal cavity losses ($j = int$).  The 
$\hat \xi_j$ and $\hat \eta$ noise operators satisfy:
\begin{align}
	[\hat\xi_j(t),\hat\xi_k^\dagger(t')]&  = \delta_{j,k} \delta(t - t'),\\
 	[\hat\eta(t),\hat\eta^\dagger(t')] & =  \delta(t - t'),\\
	[\hat\xi_j(t),\hat\eta^\dagger(t')] & =  0,\\
	\langle\hat\xi_j^\dagger(t)\hat\xi_j(t')\rangle & =  n_{c,j} \delta(t-t'),\\
	\langle\hat\eta^\dagger(t)\hat\eta(t')\rangle & =  n_{m0}^T\delta(t-t'),
\end{align}
where $n_{c,j}$ and $n_{m0}^T$ are the thermal occupation factors associated with the cavity baths and mechanical bath, respectively.  Finally, the total cavity occupancy is defined as:
\begin{equation}	
	n_c = \sum_{j=L, R, int} \frac{\kappa_j}{\kappa} n_{c,j}
\end{equation}

\subsection{Balanced detuned two-tone measurement}\label{DTT}

The first experimental measurement is performed using two balanced tones detuned by $\pm(\omega_m+\Delta)$ from the microwave resonance, where we set $\Delta=\textrm{5kHz}\ll\kappa$. 
In addition, a third cooling tone is applied at $\omega_c-\omega_m-\delta\omega_{cool}$, whose sole purpose is to increase the mechanical linewidth from $\Gamma_{m0}$ to 
$\Gamma_{m} = 100 \textrm{ Hz}$ and reduce the mechanical thermal occupancy.  We set $\delta \omega_{cool} = \textrm{35kHz}  \gg \Gamma_{m}$, which implies that the cooling tone acts independently from the measurement tones.

The average microwave field amplitude takes the form (in our interaction picture)
\begin{equation}
	\alpha(t) = 
			\bar{a}_{cool} e^{i (\omega_m + \delta \omega_{cool}) } 
		+	\bar{a}_{pump} \cos( (\omega_m + \Delta) t) .
\end{equation}

We define $n_{p}=\left|\bar{a}_{pump}\right|^{2}/4$ and $\Gamma_{opt}=\frac{4g_0^{2}}{\kappa}n_{p}$.

We solve the Heisenberg equations of motion, Eq.~\ref{heisenberg.theory1} and Eq.~\ref{heisenberg.theory2}, and calculate symmetrized noise spectral densities defined as,
\begin{equation}
S_{q}(\omega) \equiv \frac{1}{2} \int_{-\infty}^{\infty}\langle\hat q(t)\hat q^\dagger(0) + \hat q^\dagger(0)\hat q(t)\rangle e^{i\omega t}dt,
\end{equation}
for an operator $\hat q$.

Ignoring small bad-cavity corrections (which are small as $(\kappa / \omega_m)^2$), the noise spectral density of mechanical motion is found to be
\begin{align}
S_x(\omega)=\frac{x_{zp}^2\Gamma_{m}}{\omega^2+\Gamma_{m}^2/4}(1+2n_m^T+2n_{ba}),
\end{align}
where $\Gamma_{m}$ is the enhanced mechanical linewidth (i.e.~it includes the optical damping associated with the cooling tone), $n_m^T$ is the cavity-cooled mechanical thermal occupancy, and $n_{ba}= (\Gamma_{opt}/\Gamma_m ) (2n_c+1)$ is the additional heating due to the measurement tones.
It thus follows that $\langle x^2 \rangle = x_{zpt}^2 (1 + 2 \bar{n}_m)$, where the effective mechanical thermal occupancy is $\bar{n}_{m}=n_{m}^{T}+n_{ba}$ .

The output spectrum of the up-converted sideband (centered at $\omega_c-\Delta$ in the lab frame) is,
\begin{align}
	S_{R}^{red}\left(\omega + \Delta \right)	=
			\frac{1}{2}  +  n_{c,R} + \frac{4\kappa_{R}}{\kappa} ( n_{c} - n_{c,R} ) 
			+\frac{\kappa_{R}}{\kappa}\frac{\Gamma_{opt}\Gamma_{m}}{\omega^{2}+\Gamma_{m}^{2}/{4}}
	\left(\bar{n}_{m} + n_{c,R}  -2n_{c}\right),
\end{align}
where we have assumed $|\omega + \Delta| \ll \kappa$.  Similarly, the output spectrum of the down-converted sideband (centered at $\omega_c+\Delta$ in the lab frame) is,
\begin{align}
	S_{R}^{blue}\left(\omega - \Delta \right)	=
			\frac{1}{2}  +  n_{c,R} + \frac{4\kappa_{R}}{\kappa} ( n_{c} - n_{c,R} ) 
			+	\frac{\kappa_{R}}{\kappa}\frac{\Gamma_{opt}\Gamma_m}{\omega^{2}+\Gamma_{m}^{2}/{4}}\left(\bar{n}_{m}+1- n_{c,R} +2n_{c}\right),
\end{align}
where we take $|\omega - \Delta| \ll \kappa$

We see that the total weight of the two mechanically-induced sidebands in the output spectrum have information on both the mechanical occupancy $\bar{n}_m$ as well as the cavity thermal occupancy $n_c$ and the amount of thermal cavity noise incident from the right port, $n_{c,R}$.  Note that the cavity thermal noises affect the sidebands in an asymmetric manner.  By taking the average of the two sideband spectra and then fitting to a Lorentzian, we can extract $\bar{n}_m$ without needing to independently measure $n_c$, $n_{c,R}$.  

\subsection{Back-action evading measurement}

Our experiment differs from previous works and analysis \cite{ClerkNJP2008} as four drive tones are used:  two principle BAE tones that allow for a BAE measurement of the $\hat{X}_1$ mechanical quadrature by monitoring the cavity output at resonance, and two weaker tones which allows one to simultaneously (and weakly) probe a second mechanical quadrature $\hat{X}_{\phi}$ using microwaves slightly detuned a frequency $\delta \ll \kappa$ from the cavity resonance.  Similar to the detuned two-tone measurements described in Sec. \ref{DTT}, we also employ a fifth cooling tone detuned from the red mechanical sideband; similar to that case, this tone only serves to increase the mechanical linewidth and reduce its thermal occupancy.   Our analysis below also includes the impact of cavity thermal noise.

The average microwave fields due to the four measurement tones and one cooling tone thus takes the form:
\begin{equation}
\alpha(t) = \bar{a}_{pump}\cos\omega_m t +  \bar{a}_{probe} e^{-i \delta t} \cos(\omega_m t+\phi)   + \bar{a}_{cool} e^{i (\omega_m + \delta \omega_{cool}) } 
\end{equation}
where $\delta$ represents the offset in the centre frequency of the probe tones versus the principle BAE measurement tones (the ``pump" tones).  
The presence of both sets of measurement tones leads to a small deviation from a perfect QND interaction Hamiltonian.  We define the interaction strengths $G_{pump} = g_0 \bar{a}_{pump}/2$,
$G_{probe} = g_0 \bar{a}_{probe}/2$.   The linearized optomechanical interaction Hamiltonian (in an interaction picture with respect to the bare cavity and mechanical Hamiltonian) takes the form
\begin{eqnarray}
	\hat{H}_{\rm int} = 
		\frac{1}{x_{zp}} \left[
			 G_{pump} (\hat{d} + \hat{d}^\dagger) \hat{X}_1 +
			 G_{probe} (\hat{d} e^{-i \delta t} + \hat{d}^\dagger e^{i \delta t}) \left( \cos \phi \hat{X}_1 - \sin \phi \hat{X}_2 \right) \right]
		\label{eq:HInt2BAE}
\end{eqnarray}			
where we have neglected writing the interaction with the cooling tone.  The quadrature operators are defined as:
\begin{eqnarray}
	\hat{X}_1 & = &
		x_{zp} \left(  \hat{b} + \hat{b}^\dagger \right)  \, , \, 
	\hat{X}_2  = 
		-i x_{zp} \left(  \hat{b} -  \hat{b}^\dagger \right) \\			
	\hat{X}_\phi & = &
		x_{zp} \left(  \hat{b} e^{i \phi} +  \hat{b}^\dagger e^{-i \phi} \right) 
			= \cos \phi \hat{X}_1 - \sin \phi \hat{X}_2 
\end{eqnarray}			
The $G_{probe}$ term in Eq.~(\ref{eq:HInt2BAE}) describes the measurement of the $\hat{X}_\phi$ quadrature by the cavity field at $\omega = \delta$ (in the interaction picture); it also induces a backaction on the $\hat{X}_1$ quadrature measured by the main measurement tones (the ``pump" tones).  By taking $\bar{a}_{probe}$ sufficiently small and $|\delta| \gg \Gamma_m$, this additional backaction can be minimal, as we show below.

We first establish that the cavity output spectrum near $\omega = 0$ (near $\omega = \delta$) has information on the mechanical $X_1$ ($X_\phi$) quadrature.  This follows immediately from the Fourier-transformed equation of motion for the cavity field:
\begin{eqnarray}
	\hat{d}[\omega] = \frac{1}{-i \omega + \kappa/2} \left(
		-i \frac{G_{pump}}{x_{zp}} \hat{X}_1[\omega]  -i \frac{G_{probe}}{x_{zp}} \hat{X}_\phi[\omega - \delta] - \sum_{j=L,R,int} \sqrt{\kappa_{j}} \hat{d}_{j,in} \right)
\end{eqnarray}
As the mechanical quadrature operators $\hat{X}_j[\omega]$ are only appreciably non-zero for $|\omega| \lesssim \Gamma_m$, and as $\delta \gg \Gamma_m$, we see that $\hat{d}[\omega \simeq 0]$ only contains information about $X_1$, while $\hat{d}[\omega \simeq \delta]$ only contains information on $X_\phi$.  Using the standard input-output relation $\hat{d}_{j, out}(t) = \hat{d}_{j, in}(t) +
\sqrt{\kappa_j} \hat{d}(t)$, we see that this statement also carries over to the measured output field leaving the right port, $\hat{d}_{R,out}[\omega]$. 

We now turn to the additional backaction effect associated with the probe tones.  We solve Eqs. \eqref{heisenberg.theory1} and \eqref{heisenberg.theory2} to leading order in the probe-tone amplitudes.  As the Hamiltonian deviates from the perfect QND limit, there is now a weak additional optical damping of the mechanical quadratures. In the limit of interest 
($|\delta| \gg \Gamma_m$), these additional damping terms $\Gamma_{extra}$ scale as:
\begin{eqnarray}
	 \Gamma_{extra} \propto  \Gamma_m \frac{ \Gamma_{opt}^2 }{\delta^2} \left( \frac{ \bar{a}_{probe} }{ \bar{a}_{pump} } \right)^2 
\end{eqnarray}
For the parameters of our measurement, this additional damping rate is approximately $10^{-4}$ times smaller than $\Gamma_m$, and thus plays no role.  

Ignoring this non-QND damping effect, we can easily calculate the noise spectra of the mechanical quadratures, again considering the effects of the probe tones to leading order.  Consider first the $\hat{X}_2$ quadrature, which is heating by the backaction of the main (pump-tone) quadrature measurement.  We find
\begin{align}
	S_{X_2}(\omega)/x_{zp}^2 &= \frac{\Gamma_m}{\omega^2+(\Gamma_m/2)^2} \left(1+2n_m^T+2n_{ba}^{BAE}\right),
\end{align}
Here, 
\begin{equation}
	n^{BAE}_{ba}=\left(\frac{2g_0^2}{\kappa\Gamma_m}\right)\left|\bar{a}_{pump}\right|^2 (2n_c+1),
\end{equation}	
is the expected backaction heating of the $\hat{X}_2$ quadrature by the main measurement tones; note this heating is enhanced by the presence of cavity thermal noise.  As this is the dominant backaction heating of $\hat{X}_2$, we have neglected much smaller contributions associated with the deviations from the good cavity limit, and with the presence of the probe tones.

In a similar fashion, the quadrature fluctuations of the $X_1$ quadrature take the form:
\begin{align}
S_{X_1}(\omega)/x_{zp}^2 &= \frac{\Gamma_m}{\omega^2+(\Gamma_m/2)^2} \left(1+2n_m^T+2n_{bad} + 2n_{extra}  \right).
\end{align}
In this case, there is no heating by the main measurement tones in the good cavity limit; hence we retain corrections due to the finite sideband resolution and to due to the probe tones.  
$n_{bad}= \frac{1}{32}\left(\frac{\kappa}{\omega_m}\right)^2 n^{BAE}_{ba}$ describes the small of the $X_1$ quadrature due to the finite sideband resolution\cite{ClerkNJP2008}.  In contrast, $n_{extra}$ represents 
additional backaction on $X_1$ due to the probe tones.  One finds:
\begin{equation}
	n_{extra} \propto n^{BAE}_{ba} \sin^2 \phi \left( \frac{\bar{a}_{probe} }{\bar{a}_{pump}}  \right)^2
\end{equation}
Again, given the weak probe tone power used in our experiments, the additional heating is negligible (e.g. for the results in Fig.~3b of the main text,  $n_{extra}$ is at most $ \sim 0.3$, smaller than the resolution in the $X_1$ measurement).

A last effect of the probe backaction is the generation of small (classical) noise correlations between $X_1$ and $X_2$, i.e. the symmetrized noise correlator $S_{X_1 X_2}[\omega]$ will not be zero.
As this correlation arises solely from the probe backaction, it will be proportional to $( \cos \phi \sin \phi )n_{extra}$, and is thus again negligible for our experiments; we thus do not include it in our analysis.  This then yields:
\begin{equation}
	S_{X_\phi X_\phi}(\omega) = (\cos^2 \phi) \,  S_{X_1 X_1}(\omega) + (\sin^2 \phi) \,  S_{X_2 X_2}(\omega)
\end{equation}
This form (which reflects the absence of quadrature correlations) is consistent with the experimentally observed $\phi$-dependence of the $X_\phi$ noise, see. Fig. 3 of main text.

Finally, note that for both $X_1$ and $X_2$, the backaction contributions to the fluctuations have a term proportional to the cavity thermal occupancy $n_c$.  To determine this contribution experimentally, we deliberately inject additional noise quanta into the cavity (hence increasing $n_c$).  By extrapolating the dependence of the measured spectrum on this injected noise, we can determine the quantum part of the backaction.

\section{Measurement circuit}
\begin{figure}[ht]
\centering\includegraphics{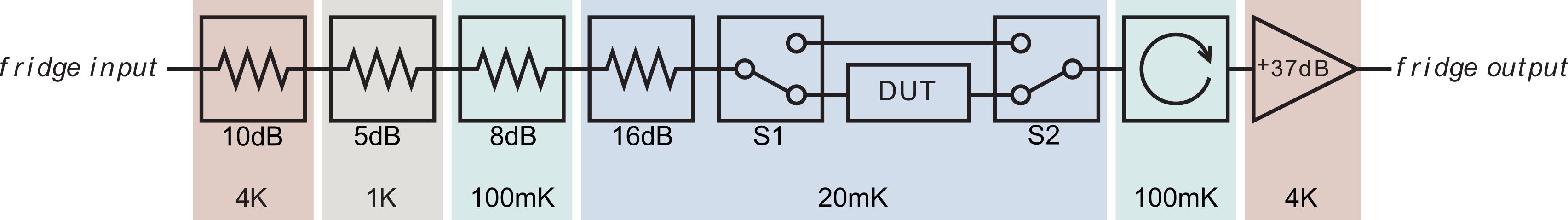}
\caption{Cryogenic part of measurement circuit}
\label{circuit}
\end{figure}

Cryogenic part of the measurement circuit is shown in Fig.\ref{circuit}. Four stages of cold Ni-Cr attenuators dissipate the room-temperature Johnson noise to keep the input microwave noise at shot-noise level. Two cryogenic microwave switches are employed to provide noise floor calibration. In the experimental mode, the device(DUT) is connected to the circuit, and in the calibration mode, a short Nb superconducting transmission line is connected. The output microwave passes through two stages of cryogenic circulators mounted at the cold plate($\approx$100mK), then gets amplified by a low-noise, high-electron mobility transistor (HEMT) amplifier at 4.2K. The circulators block the Johnson noise from 4.2K by 40dB or more. 

\begin{figure}[ht]
\centering\includegraphics{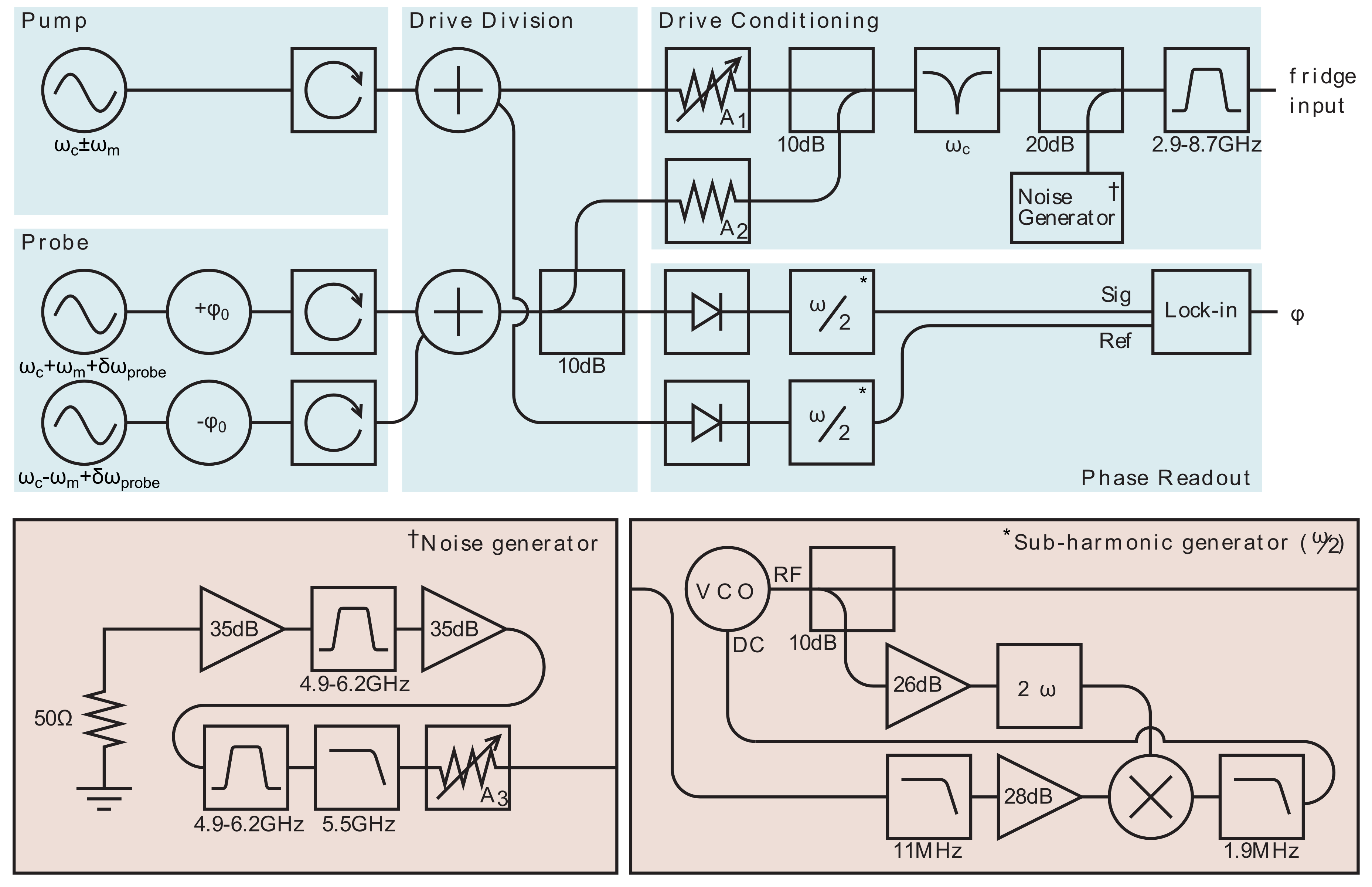}
\caption{Room temperature part of measurement circuit}
\label{doubleBAE}
\end{figure}

Room temperature part of the measurement circuit is composed of seven blocks. (Fig.\ref{doubleBAE})
\begin{enumerate}
  \item Pump: a vector source(Agilent E8267C) generates two-tone pumps (BAE or detuned).
  \item Probe: combined output from two scalar sources supply phase-locked probe tones.
  \item Drive division: fraction of microwave power is fed into Phase readout circuit, and the rest goes into drive conditioning circuit for leveling and filtering.
  \item Drive conditioning: attenuators (A1,A2), cavity filter for phase noise rejection up to 50dB, and a band-pass filter set up the input microwave into the device. A directional coupler for noise injection is also included.
  \item Phase readout: the beat tones of pumps and probes are acquired using microwave detection diodes. The beat tones go through the sub-harmonic generator circuits to generate half-frequency, and go into the signal and reference channel of lock-in, to measure the phase difference between them.
  \item Sub-harmonic generator: a standard phase-locked-loop circuit with a frequency doubler is employed to generate a output with half-frequency, which is phase-locked to the input.
  \item Noise generator: room-temperature Johnson noise is amplified and filtered around LC resonance frequency to inject excess thermal noise into the device.
\end{enumerate} 

\section{Data analysis and calibration}

\subsection{Calibration of mechanical occupation factor and number of pump photons}
Following Ref.\cite{hertzberg09th}, the motional sideband power($P_m$) normalized by the output pump power($P_{thru}$) is,

\begin{equation}
\frac{P_m}{P_{thru}}=4\left(\frac{g_0}{\kappa}\right)^2 \frac{k_BT_m}{\hbar\omega_m},
\end{equation}

\noindent where $g_0$: single photon optomechanical coupling($=(\partial\omega_c/\partial x)x_{zp}$), $k_B$: Boltzmann constant, and $T_m$: mechanical mode temperature. At various calibrated temperatures $T_m$, which are controlled by dilution refrigerator, a weak pump at $\omega_c-\omega_m$ is applied and its motional sideband power is measured through spectrum analyzer. $P_{thru}$ for the single pump tone(``red'') at $\omega_c-\omega_m$, is kept small to minimize the back-action damping less than 1 Hz. The data fit yields, $n_m=k_BT_m/(\hbar \omega_m)=(9.92 \pm 0.16)\times10^8\cdot(P_m/P_{thru})$ for the mechanical occupation factor $n_m$(Fig.1c, main text). This is the calibration for $n_m$ used in this work.

The mechanical occupation factor calibration also yields the single photon optomechanical coupling $g_0=(13.8\pm0.2)$ Hz. We measure back-action damping in a driven mechanical response set-up with a single red tone at $\omega_c-\omega_m$ and a sweeping probe tone near $\omega_c$. The probe output shows Lorentzian absorption spectrum due to the electro-mechanical coupling, similar to electromagnetic transparency\cite{Safavi:2011}. The back-action damping results in the linewidth broadening of mechanical spectrum by $\Gamma_{opt}$ as,

\begin{equation}
\Gamma_{opt} =  \frac{4g_0^2}{\kappa}n_p.
\end{equation}

Since the output pump power $P_{thru}$ is proportional to the average pump photon number $n_p$, we get the factor $\beta=n_p/P_{thru}=(\Gamma_{opt}/P_{thru})/(4g_0^2/\kappa)=(2.25\pm0.07)\times10^{11} (\textrm{W}^{-1})$, which serves as the calibration for average pump photon number.

\subsection{Measurement of microwave occupation factor from noise spectral density}

The Langevin equation of the microwave resonator is

\begin{align}
\dot{\hat{d}}\left(t\right)=-\left(i\omega_{c}+\frac{\kappa}{2}\right)\hat{d}\left(t\right)-\sqrt{\kappa}\hat{\xi}\left(t\right),
\end{align}
where $\hat{\xi}\left(t\right)\equiv\underset{i=L,R,int}{\sum}\sqrt{\frac{\kappa_{i}}{\kappa}}\hat{\xi}_{i}\left(t\right)$, following definitions in section \ref{theory}.

In frequency domain, 

\begin{equation}
\hat{d}\left(\omega\right)=-\sqrt{\kappa}\chi_{c}\left(\omega\right)\hat{\xi}\left(\omega\right),
\end{equation}

\noindent where $\chi_{c}\left(\omega\right)=\frac{1}{\frac{\kappa}{2}-i\left(\omega-\omega_{c}\right)}$.

According to the standard input output theory, the output field of the microwave resonator is given by

\begin{equation}
\hat{\xi}_{R,out}\left(\omega\right)=\hat{\xi}_{R}\left(\omega\right)+\sqrt{\kappa_{R}}\hat{d}\left(\omega\right)=\hat{\xi}_{R}\left(\omega\right)-\sqrt{\kappa_{R}\kappa}\chi_{c}\left(\omega\right)\hat{\xi}\left(\omega\right).
\end{equation}

We measure the symmetrized spectral density of microwave field:
\begin{align}
\bar{S}_{R}\left(\omega\right) & =  \frac{1}{2}\int dt\, e^{i\omega t}\left\langle \left\{ \hat{\xi}_{R,out}\left(t\right),\hat{\xi}_{R,out}^{\dagger}\left(0\right)\right\} \right\rangle \\
 & =  \frac{\kappa_{R}\kappa}{\left(\frac{\kappa}{2}\right)^{2}+\left(\omega-\omega_{c}\right)^{2}}\left(n_{c}-n_{c,R}\right)+n_{c,R}+\frac{1}{2},\label{eq.sr}
\end{align}
where $n_c$ is defined in section \ref{theory}.

To convert this spectral density into the voltage noise density measured at the spectrum analyzer, we set a conversion factor $\alpha$ between the measured voltage noise $S_V$ and $\frac{\kappa}{4\kappa_R}\cdot\bar{S}_{R}$ as,
\begin{align}
S_V\left(\omega\right)=\frac{1}{\alpha}\cdot\frac{\kappa}{4\kappa_R}\cdot\bar{S}_{R}\left(\omega\right)+S_V^{HEMT},\label{eq.sv}
\end{align}
where $S_V$ also contains the noise from HEMT amplifier ($S_V^{HEMT}$).

We measure $S_V$ at the resonance, and call the measured noise density as $\eta\left(=S_V\left(\omega_c\right)\right)$. From Eq.\ref{eq.sr} and Eq.\ref{eq.sv},
\begin{align}
\alpha\eta=n_c-n_{c,R}+\left(\frac{\kappa}{4\kappa_R}\right)n_{c,R}+\frac{\kappa}{8\kappa_R}+\alpha S_V^{HEMT}.\label{eq.eta}
\end{align}

Since the circulators reduces 4K noise by more than 40dB, the thermal noise from the output port, $n_{c,R}$, is dominated by the terminations on the isolated ports of circulators at the cold plate. To extract $n_c$, we calibrate the background noise $\eta_0$ by measuring the noise floor after switching from the device to a short transmission line using cryogenic microwave switches. In order to guarantee equal HEMT gain and noise between the experiment and calbration, we keep the HEMT output power of measurement tones constant. Now in $\eta_0$ measurement, the device contribution (the first and second term in Eq.\ref{eq.eta}) is no longer present. In addition, the thermal noise from the output port becomes absorbed by the matched input line attenuators without being reflected, and the third term in Eq.\ref{eq.eta} is also absent. Therefore,
\begin{align}
\alpha\eta_0=\frac{\kappa}{8\kappa_R}+\alpha S_V^{HEMT}.
\end{align}

By taking the difference $\Delta\eta=\eta-\eta_0$, $n_c$ reads
\begin{align}
n_c = \alpha\Delta\eta+ \left(\frac{4\kappa_R-\kappa}{4\kappa_R}\right)n_{c,R}.
\end{align}
The normalized back-action in $X_2$ quadrature (see Fig.3d) is
\begin{align}
\frac{\left<X_2^2\right>_{ba}}{\left<X_2^2\right>_{qba}} &= 2n_c+1\\
&=2\alpha\Delta\eta + 1 + \left(\frac{4\kappa_R-\kappa}{2\kappa_R}\right)n_{c,R},\label{eq.2nc}.
\end{align}
From Fig.3d, $\alpha=0.22\pm0.02$ (aW/Hz)$^{-1}$. We see $n_{c,R}$ modifies the measured intercept from the quantum back-action term, ``+1''. 

\begin{figure}[ht]
\centering\includegraphics{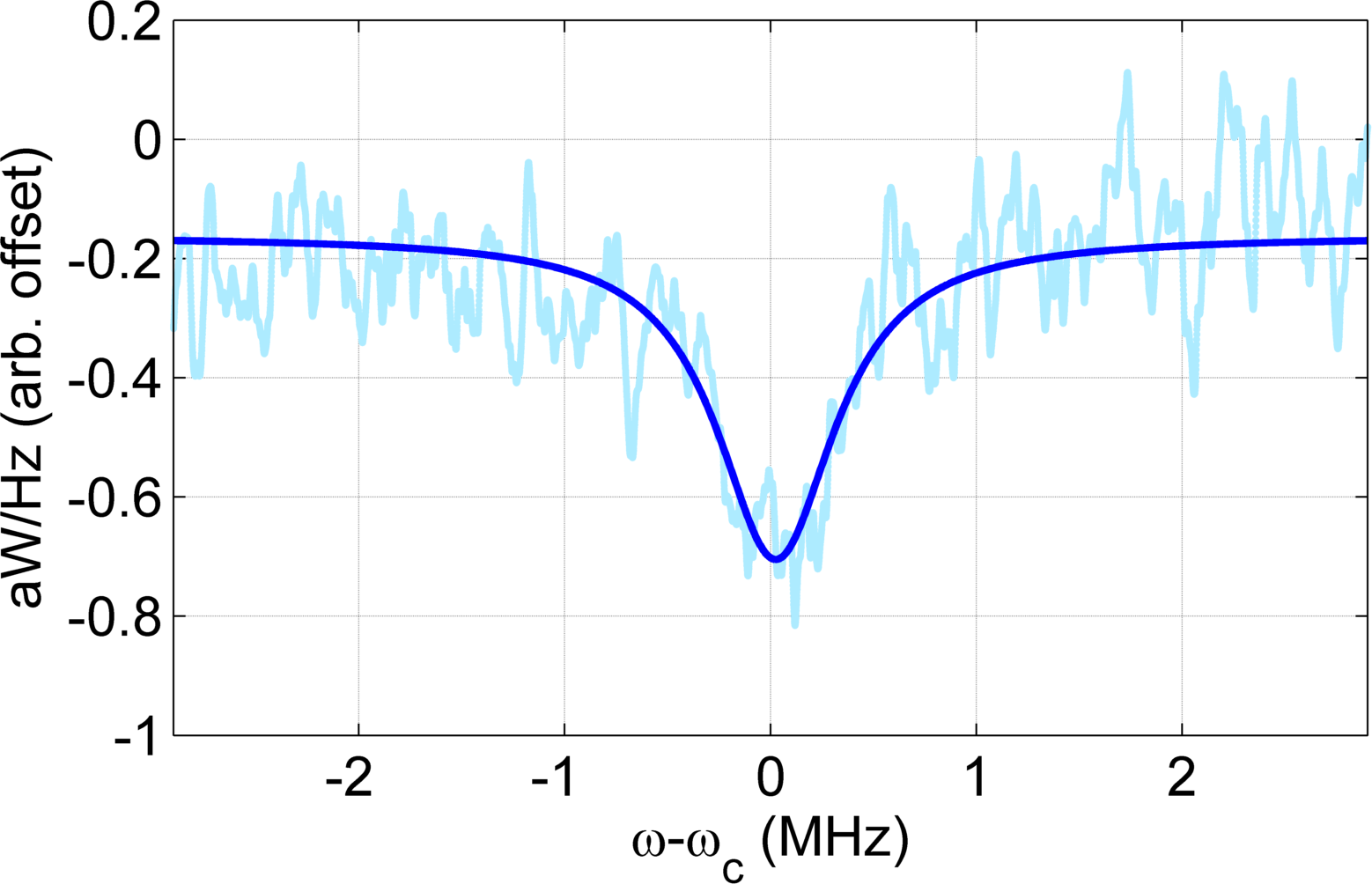}
\caption{Noise from LC resonator with no pump and no injected noise}
\label{cavnoise}
\end{figure}

We measure $n_{c,R}$ by probing the noise spectrum around $\omega_c$ when the noise from other sources are negligible than from the output port, i.e. when no pumps and no injected noise is present(Fig.\ref{cavnoise}). In this case, $n_c=\frac{\kappa_R}{\kappa}n_{c,R}$ and from Eq.\ref{eq.sr} and Eq.\ref{eq.sv},
\begin{align}
S_V\left(\omega\right)=\frac{1}{\alpha}\cdot\left(\frac{\kappa^2}{\kappa^2+4\left(\omega-\omega_{c}\right)^2}\left(\frac{\kappa_R}{\kappa}-1\right)n_{c,R}+\left(\frac{\kappa}{4\kappa_R}\right)n_{c,R}+\frac{\kappa}{8\kappa_R}\right)+S_V^{HEMT}
\end{align}
and we estimate $n_{c,R}\approx0.2$ from the measured area under Lorentzian, with $\kappa=2\pi\times0.86$ MHz and $\kappa_R=2\pi\times0.45$ MHz. ($\kappa_R$ is measured at 300mK in a separate set-up.) This $n_{c,R}$ would add a minor correction about 0.2 to the quantum back-action, ``+1'', and it is consistent to the intercept of $1.1\pm0.1$ from the fit in Fig.3d.

\subsection{BAE measurement}
Fine balancing of two tones is necessary for precise BAE measurement. For instance, at high $n_p$, where $\Gamma_{opt}\cong1$kHz, 1\% imbalance in power results in modification of $\Gamma_m$ by 10Hz. We achieve the fine tuning by matching $\Gamma_m$ to $\Gamma_{m,init}$ with 0.01dB level of tuning in two-tone power ratio, where $\Gamma_{m,init}$ is the mechanical linewidth measured before applying two tones. The procedure we follow is:
\begin{enumerate}
  \item Apply cooling tone($n_p\approx10^5$) and take its upper-sideband to measure $\Gamma_{m,init}$.
  \item Apply two detuned tones at $\omega_c\pm(\omega_m+5\textrm{kHz})$(-:red,+:blue).
  \item Measure sidebands of two tones and acquire $\Gamma_m$
  \item Tune the ratio between two tones to match $\Gamma_m=\Gamma_{m,init}$ within $\pm5$ Hz.
\end{enumerate}

\begin{figure}[ht]
\centering\includegraphics{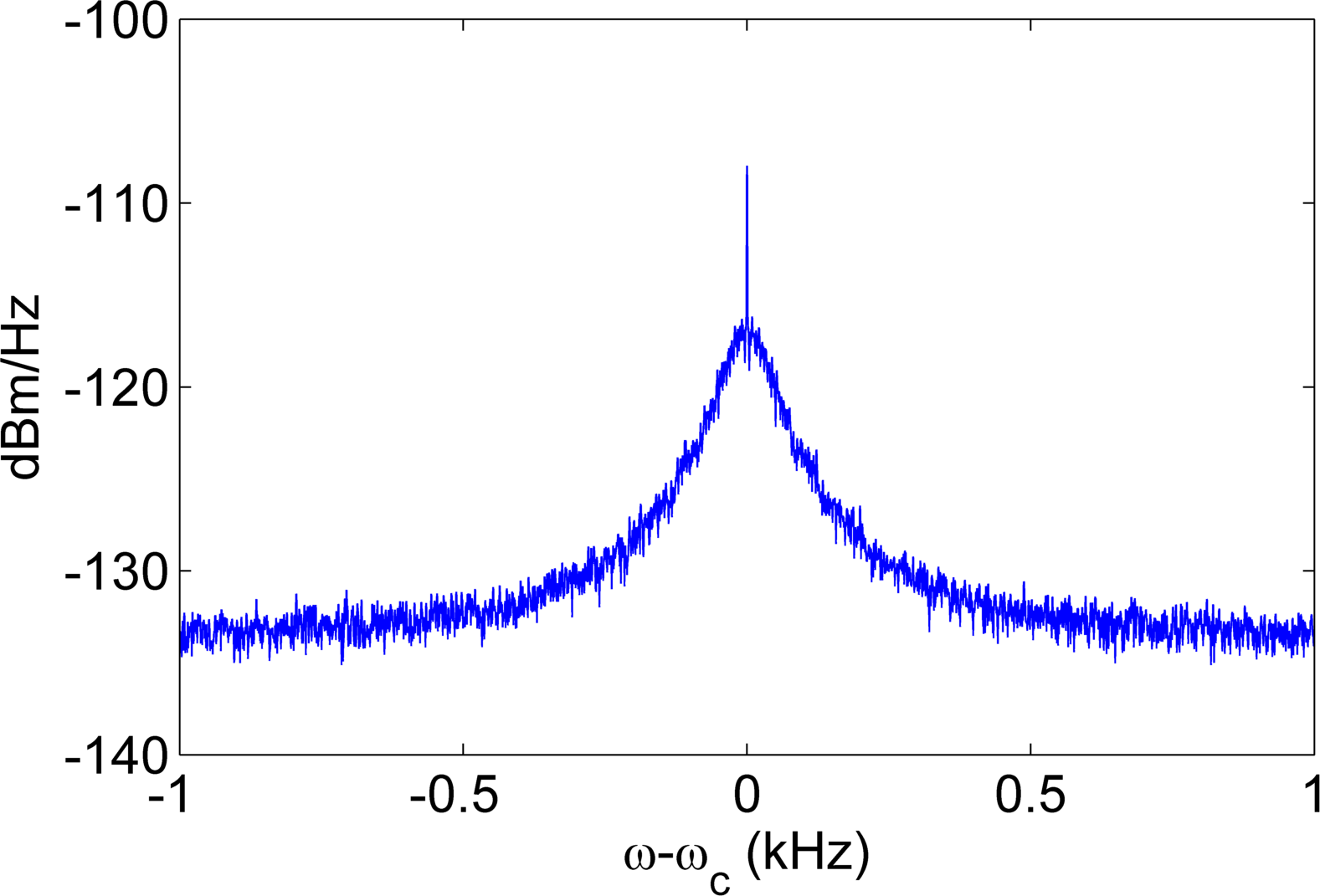}
\caption{Example of a mechanical spectrum in BAE measurement}
\label{BAEspec}
\end{figure}

BAE measurement is done by using the powers found from the procedure above, and setting two-tone frequencies at $\omega_c\pm\omega_m$. We take the motional sideband spectra using an FFT spectrum analyzer(Agilent N9520A). Figure S1 is an example from a BAE measurement. The vector source generating BAE pump tones has a small spur at the middle of two tones, overlapping with the motional sideband which can be seen as a spike in the middle in Fig.\ref{BAEspec}. This makes it necessary to mask a narrow section of spectra, 5Hz wide, for proper fitting to Lorentzian. Compared to the linewidth of mechanical resonance($\Gamma_m\simeq100$ Hz), the masked section is negligible.

\bibliographystyle{plainnat}